\documentclass[twocolumn]{aastex62}
\usepackage{microtype}
\usepackage{amsmath}
\usepackage{booktabs}
\usepackage{xcolor}
\newcommand{\sci}[2]{{#1}\times10^{#2}}
\newcommand{\Msun}{\,\mathrm{M}_\odot}
\newcommand{\Rsun}{\,\mathrm{R}_\odot}
\newcommand{\Lsun}{\,\mathrm{L}_\odot}
\newcommand{\uHz}{\,\mu\mathrm{Hz}}

\newcommand{\s}{\,\mathrm{s}}
\newcommand{\Gyr}{\,\mathrm{Gyr}}
\newcommand{\K}{\,\mathrm{K}}
\newcommand{\degrees}{{^\circ}}
\newcommand{\FeH}{[\mathrm{Fe}/\mathrm{H}]}
\newcommand{\Teff}{T_\mathrm{eff}}
\newcommand{\numax}{\nu_\mathrm{max}}
\newcommand{\eq}[1]{\begin{equation} #1 \end{equation}}
\newcommand{\st}[1]{_\mathrm{#1}}

\begin{document}

\title{A synthetic sample of short-cadence solar-like oscillators for TESS}

\author[0000-0002-4773-1017]{Warrick H. Ball}
\affiliation{School of Physics and Astronomy, University of Birmingham, Edgbaston, Birmingham B15 2TT, United Kingdom}
\affiliation{Stellar Astrophysics Centre, Department of Physics and Astronomy, Aarhus University, Ny Munkegade 120, DK-8000 Aarhus C, Denmark}

\author[0000-0002-5714-8618]{William J. Chaplin}
\affiliation{School of Physics and Astronomy, University of Birmingham, Edgbaston, Birmingham B15 2TT, United Kingdom}
\affiliation{Stellar Astrophysics Centre, Department of Physics and Astronomy, Aarhus University, Ny Munkegade 120, DK-8000 Aarhus C, Denmark}

\author[0000-0002-0786-7307]{Mathew Schofield}
\affiliation{School of Physics and Astronomy, University of Birmingham, Edgbaston, Birmingham B15 2TT, United Kingdom}
\affiliation{Stellar Astrophysics Centre, Department of Physics and Astronomy, Aarhus University, Ny Munkegade 120, DK-8000 Aarhus C, Denmark}

\author[0000-0001-5998-8533]{Andrea Miglio}
\affiliation{School of Physics and Astronomy, University of Birmingham, Edgbaston, Birmingham B15 2TT, United Kingdom}
\affiliation{Stellar Astrophysics Centre, Department of Physics and Astronomy, Aarhus University, Ny Munkegade 120, DK-8000 Aarhus C, Denmark}

\author{Diego Bossini}
\affiliation{INAF-Osservatorio Astronomico di Padova, Vicolo dell'Osservatorio 5, I-35122 Padova, Italy}

\author[0000-0002-4290-7351]{Guy R. Davies}
\affiliation{School of Physics and Astronomy, University of Birmingham, Edgbaston, Birmingham B15 2TT, United Kingdom}
\affiliation{Stellar Astrophysics Centre, Department of Physics and Astronomy, Aarhus University, Ny Munkegade 120, DK-8000 Aarhus C, Denmark}

\author[0000-0002-6301-3269]{L{\'e}o Girardi}
\affiliation{INAF-Osservatorio Astronomico di Padova, Vicolo dell'Osservatorio 5, I-35122 Padova, Italy}

\correspondingauthor{Warrick H. Ball}
\email{W.H.Ball@bham.ac.uk}

\begin{abstract}
  NASA's \emph{Transiting Exoplanet Survey Satellite} (TESS) has begun
  a two-year survey of most of the sky, which will include lightcurves
  for thousands of solar-like oscillators sampled at a cadence of two
  minutes.  To prepare for this steady stream of data, we present a
  mock catalogue of lightcurves, designed to realistically mimic the
  properties of the TESS sample.  In the process, we also present the
  first public release of the asteroFLAG Artificial Dataset Generator,
  which simulates lightcurves of solar-like oscillators based on input
  mode properties.  The targets are drawn from a simulation of the
  Milky Way's populations and are selected in the same way as TESS's
  true Asteroseismic Target List.  The lightcurves are produced by
  combining stellar models, pulsation calculations and semi-empirical
  models of solar-like oscillators.  We describe the details of the
  catalogue and provide several examples.  We provide pristine
    lightcurves to which noise can be added
  easily.  This mock catalogue will be valuable in testing
  asteroseismology pipelines for TESS and our methods can be applied
  in preparation and planning for other observatories and observing
  campaigns.
\end{abstract}


\section{Introduction}

The study of stellar oscillations---\emph{asteroseismology}---has
undergone a revolution, driven by space-based photometric observations
from COROT \citep{corot}, \textit{Kepler} \citep{kepler} and K2
\citep{k2}.  In particular, space-based photometry has provided data
of unprecedented quality for solar-like oscillators, whose
low-amplitude oscillations had previously been notoriously difficult
to observe.

NASA's \emph{Transiting Exoplanet Survey Satellite}
\citep[TESS,][]{tess} will extend this new era.  Like \emph{Kepler},
TESS is chiefly an exoplanet survey mission but its continuous,
high-cadence observations are also suited to the study of stellar
oscillations.  TESS will observe most of the sky in roughly month-long
sectors covering four $24\degrees\times24\degrees$ areas from the
ecliptic poles to near the ecliptic plane.  The mission will produce
full-frame images (FFIs) every 30 minutes as well as light curves for
a selection of targets sampled at a short cadence of two minutes,
  which is necessary for the seismology of cool main-sequence and subgiant stars.
Once reduced to lightcurves, the FFIs will also allow asteroseismic
analysis but here we restrict our attention to short-cadence targets.
The satellite was launched on 2018 April 18 and began science operations
on 2018 July 25.  The first data release is expected
about six months after science operations
began \citep{tess} i.e.~late January 2019.
Each month of short-cadence data is expected to include hundreds of
stars in which solar-like oscillations will be detected.

In preparation for this rapid flow of data, we present here a mock
catalogue of TESS lightcurves for a sample of
solar-like oscillators observed at short cadence.  The targets have been
  selected from a synthetic Milky Way population by the same method
as the real Asteroseismic Target List  (ATL, Schofield et al., in prep.) of the TESS
Asteroseismic Science
Consortium \citep[TASC,][]{tasc36}.\footnote{\url{https://tasoc.dk/docs/SAC_TESS_0003_6.pdf}}
These synthetic lightcurves will be used
to test parameter extraction pipelines (and potentially model-fitting
pipelines) with known physical parameters.  Although some important
quantities (e.g.~rotation rates) cannot currently be predicted a
priori, the rich phenomenology of solar-like oscillators derived from
previous missions allows us to generate realistic lightcurves using
empirical methods.  These lightcurves are also provided in a simple
format so that they can be supplemented with other signals, like
transiting planets or systematic effects.

We first present our method for producing stellar models for a
sample that mimics the ATL (Section~\ref{s:sample}) followed by
the inputs and methods by which we computed
lightcurves for each star in that sample
(Section~\ref{s:lightcurves}).  We then describe the structure of our
model catalogue and present several example results
(Section~\ref{s:catalogue}) before discussing shortcomings and
potential future applications of our methods
(Section~\ref{s:discussion}). We close our presentation with a
  brief conclusion (Section~\ref{s:conclusions}).

\section{Methods}
\label{s:sample}

\subsection{Stellar models}

All of the stellar models used in this work were computed using
Modules for Experiments in Stellar Astrophysics\footnote{\url{http://mesa.sourceforge.net}} (\textsc{MESA}),
revision 7385 \citep{paxton2011,paxton2013}.  The stellar model grids
used in the Galaxy simulation (see Sec.~\ref{ss:galsim}) are the same
as described by \citet{rodrigues2017} and the same inputs were used to
recreate individual stellar model profiles at the interpolated
parameter values (see Sec.~\ref{ss:modelparams}).  Full details are
given by \citet{rodrigues2017} but we give the main parameters again
here.

The models use the solar metal mixture of \citet{grevesse1993}, with
solar metal and helium abundances $Z_\odot=0.01756$ and
$Y_\odot=0.26618$.  Stellar models at other metallicities follow the
enrichment law $Y=0.2485+1.007\times Z$.  The atmospheric model is that
of \citet{krishna1966}, which gives a solar-calibrated mixing length
parameter $\alpha\st{MLT}=1.9657$.  Opacities are taken from the
OPAL tables \citep{iglesias1996} at high temperatures
($\log_{10}(T/\K)\geq4.1$), \citet{ferguson2005} at low temperatures
($\log_{10}(T/\K)\leq4.0$) and blended linearly between
($4.0\leq\log_{10}(T/\K)\leq4.1$).  The equation of state is the MESA
default, which is derived from the OPAL equation of state
\citet{rogers2002} in the region relevant for our stellar models.

\subsection{Galaxy simulation}
\label{ss:galsim}

We simulated the population of stars in the Milky Way using TRILEGAL
\citep{girardi2005}.  The simulation used the default parameters
  described by \citet{girardi2012}, which comprise a thin disc, thick
disc, halo and bulge.  The simulation includes a rough model for
  extinction, in which the total extinction determined by
  \citet{schlegel1998} is assumed to be caused by an exponential dust
  disc with a scale height of $110\,\mathrm{pc}$.  Bolometric
  corrections and extinction coefficients were calculated for the TESS
  bandpass in a Vega magnitude system.  The stellar models (as
described in the previous section) span masses from $0.60$ to
$2.50\Msun$ in steps of between $0.05$ and $0.20\Msun$ \citep[see
  Table 1 of ][]{rodrigues2017} and metallicities $\FeH$ from $-1.00$
to $0.50$ in steps of $0.25$.  The limits of the stellar model grid
naturally restrict our base population to those ranges.  The
population was selected to cover the whole sky down to a magnitude
limit of $12.5$ in the TESS bandpass.

\begin{figure}
  \centering
  \includegraphics[width=\columnwidth]{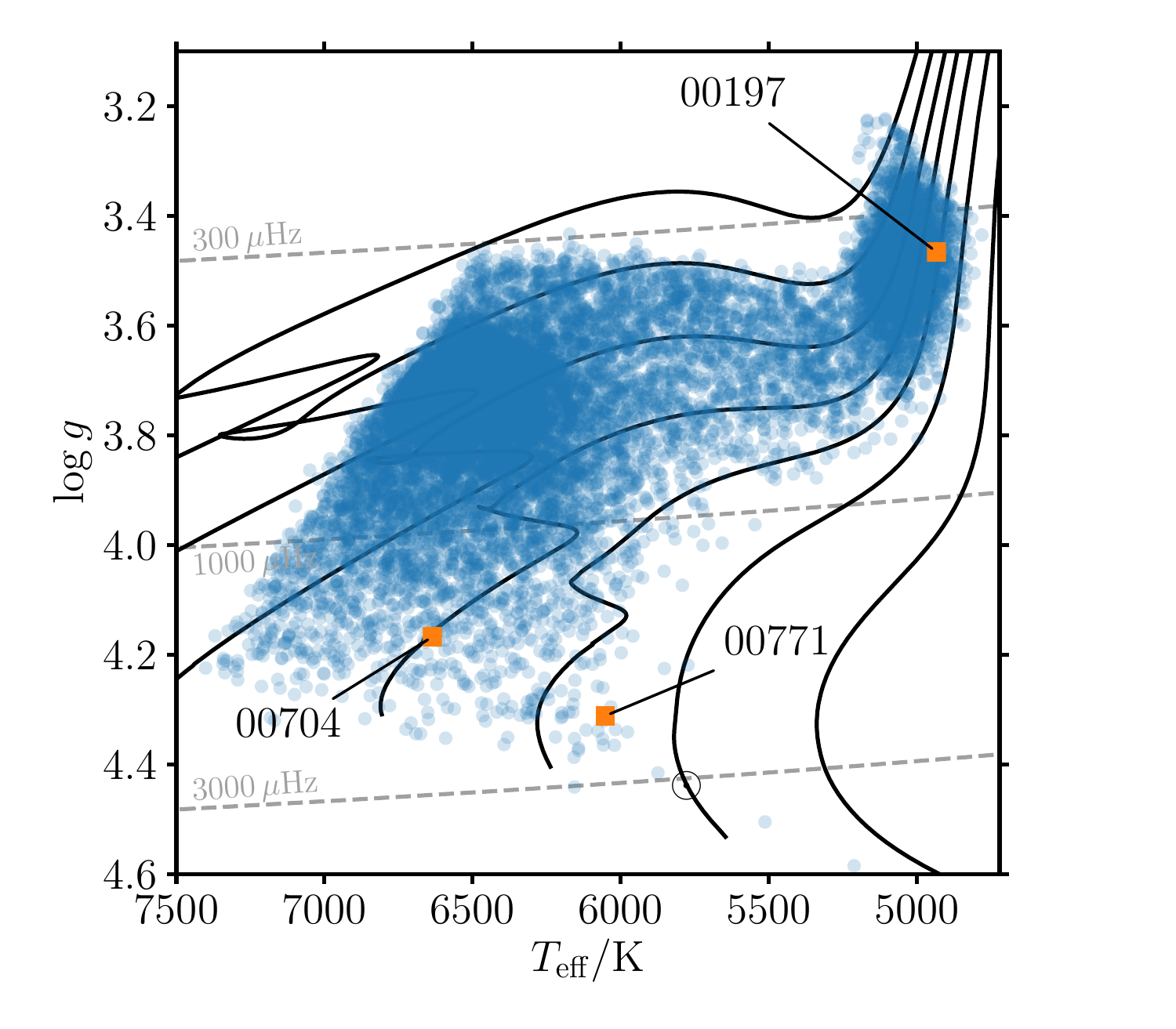}
  \caption{Kiel diagram of the stars selected by the ATL code
      from the TRILEGAL simulation.  The solid black lines are
      evolutionary tracks at solar metallicity for masses from
    $0.8\Msun$ to $2.0\Msun$ in steps of $0.2\Msun$.  Orange points
    are stars whose power spectra are shown elsewhere.  Star
      00197 appears in Figs~\ref{f:typical} and \ref{f:full}, star
      00704 in Fig.~\ref{f:Ftype} and star 00771 in
      Fig.~\ref{f:Gtype}.  The dashed grey lines, from top to bottom,
    show constant $\numax=300$, $1000$ and $3000\uHz$.
  The Sun is indicated by its usual symbol.}
  \label{f:kiel}
\end{figure}

\subsection{Target selection}

We ranked the stars in the TRILEGAL simulation by the likelihood of
detecting solar-like oscillations, as determined by the same code used
to produce the Asteroseismic Target List (ATL, Schofield et al., in prep.) for Working Groups 1
and 2 of the TESS Asteroseismic Science Consortium (TASC).  The
principles behind the code are
described by \citet{campante2016}, which determines the detection
probability by the same method as \citet{chaplin2011}, appropriately
modified for TESS using its expected noise characteristics
\citep{sullivan2015}.
In short, the code uses empirical relations to predict the
amplitude of a potential target's oscillations and compares the
star's expected noise level to determine the probability
that the oscillations will be detected.

In addition, we used the distance moduli
provided by the TRILEGAL simulation, rather than also creating mock
distances to mimic the Tycho--Gaia Astrometric Solution
\citep[TGAS, as used in][]{campante2016} or Gaia DR2
(which is being used for the final version of the ATL).

TESS's observing strategy is important because it divides the sky into
26 partially-overlapping observing sectors of varying durations.
We shall partially classify our results by these sectors.
The satellite observes each hemisphere of the sky in
sectors containing four $24\degrees\times24\degrees$ areas that
cover a strip of the sky from the ecliptic pole to near the
ecliptic plane.  Each hemisphere is observed for 13 sectors and each
sector is observed for about 27.4 days on average.
After observing in one hemisphere,
the satellite will re-orient to observe the other hemisphere.
The first sector is centred on a galactic longitude of
$315.8\degrees$,\footnote{\url{https://tess.mit.edu/observations/sector-1/}}
  which we have replicated in our mock sample.

TESS uses an orbit in 2:1 resonance with the moon, in which the
  lengths of the orbits vary within a range of a few days
  \citep{dichmann2014, dichmann2016}.  The details of the pointings
  depend on the precise orientation of the spacecraft and these are
  not known until the observations of a given sector begin.  In our
  mock sample, we have extrapolated approximate pointings and sector
  durations from perigee data provided by the TESS
  team.\footnote{\url{https://figshare.com/articles/TESS_Perigee_Times/6875525}}
  For the sector pointings, we fit the times (as Julian dates) of the
  mid-sector perigees $t\st{perigee}$ with the formula
\begin{align}
  t\st{perigee}&=2458339.922 + 27.266\,(n-1) \\
  &\qquad + 2.386\sin(2\pi(0.0937\,(n-1)-0.0683)
\end{align}
where $n$ is the sector number, from 1 to 26.  We then took the
  pointing of a given sector to be the anti-solar direction at that
  time.  For the sector durations $\Delta t$, we fit a similar formula to the
  durations between perigees at the start and end of each sector,
\eq{\Delta t/\mathrm{d}=27.276 + 1.493\sin(2\pi(0.00345\,t+0.171))}
where $t$ is the Julian date at the start of a sector. Though not
  perfectly accurate, these formulae give our mock sample a realistic
  variation in the durations and pointings of each sector all the way
  to the end of the nominal mission.

Running the ATL code on our TRILEGAL simulation data provided a
ranking for all the stars in the simulation.  To create our mock
  sample, we selected enough stars for each sector to contain at least
  1000 stars.  The mock sample contains 12731 stars and each sector
  contains between 1000 and 1263 targets (many of which appear
    in more than one sector).  In the output ATL target
  list, $99.07$ per cent of the stars are in the thin disc, $0.85$ per
  cent in the thick disc, $0.08$ per cent in the halo and none in the
  bulge.  Fig.~\ref{f:kiel} shows the stars in the Kiel diagram
    (effective temperature $\Teff$ versus surface gravity $\log g$).

Note that the ATL target selection presumably contaminates
  the sample with some number of classical pulsators, in particular
  $\gamma$ Doradus variables.  Indeed, $\gamma$ Doradus itself is one
  of the targets in the real ATL.  This is a deliberate choice to
    better sample the transition from solar-like oscillations to coherent
  pulsations and to search for potential hybrid oscillators.
  We have assumed that all the
  ATL-selected stars are solar-like oscillators and have ignored this
  contamination.  Using various estimates of the red edge of the
  $\gamma$ Doradus instability strip \citep[e.g.][]{dupret2004}, we
  estimate that as much as 20 per cent of our sample might be in the
  instability strip.  Most of these stars, however, are ranked in the
  lowest quarter of our sample.

The target selection also does not account for the binarity of
  systems, even though the TRILEGAL simulation does generate stars in
  binary systems.  The ATL selects its targets using the single star
  data and therefore selects targets that might actually be difficult
  to observe because of a companion.  For reference, TRILEGAL labels
  64.0 per cent of the selected targets as single stars, 28.0 per
  cent as primaries and 8.0 per cent as secondaries.

\subsection{Stellar model parameters}
\label{ss:modelparams}

The simulated population does not provide complete stellar models,
which are required to compute mode frequencies, so we recomputed
evolutionary tracks with the initial parameters in the TRILEGAL
simulation and stored the final models of these tracks for the
oscillation calculation.  Because TRILEGAL interpolates in a grid of
models to compute a star's observable properties, we expect some
differences in stellar properties caused by
interpolation.  We initially proceeded naively, using exactly
the stellar age given by the TRILEGAL data.  This gave differences of
up to about $5$ per cent in $\Teff$ and $\log(L/L_\odot)$.

To reduce the differences in these key properties, we instead evolved
the star until the misfit in $\Teff$ and $\log(L/L_\odot)$ reached a
minimum near the age given in the TRILEGAL data.  This improved the
mean accuracy to better than about 0.5 per cent at the cost of
introducing small discrepancies in the ages, of up to about 1 per cent
on average.  Specifically, we minimised the misfit
\begin{align}
  \chi^2
  &=\left(\frac{T\st{eff,MESA}-T\st{eff,TRI}}{150\K}\right)^2 \\
  &+\left(\frac{\log_{10}(L\st{MESA}/L_\odot)-\log_{10}(L\st{TRI}/L_\odot)}{0.03}\right)^2
\end{align}
where the subscripts $\textsc{MESA}$ and $\textsc{TRI}$ indicate
quantities from the recomputed MESA model or the TRILEGAL data.
The uncertainties were chosen to balance the quality of the match
  between the luminosity $L$ and the effective temperature $\Teff$.
  This is unimportant for most stars but we chose these values
  to avoid local minima of $\chi^2$ in stars on or just beyond
  the blue hook.

Like \citet{rodrigues2017}, we evolved our models starting from the
pre-main-sequence.  MESA sometimes fails to converge on initial models
when using atmospheric $T(\tau)$ relations so \citet{rodrigues2017}
used different values of the initial central temperature $T_c$ for
different evolutionary tracks.  When recomputing models at
interpolated parameter values from the TRILEGAL output, we also
interpolated $T_c$ linearly as a function of mass $M$ and metallicity
$Z$.  This still led to some runs failing to converge on an initial
model.  In these cases, we increased the initial central temperature by
$1000\K$ at a time until an initial model converged and the run could
proceed.  About 4 per cent of all our models require this step, and
just over half of those require just one change to $T_c$.

\begin{figure}
  \centering
  \includegraphics[width=\columnwidth]{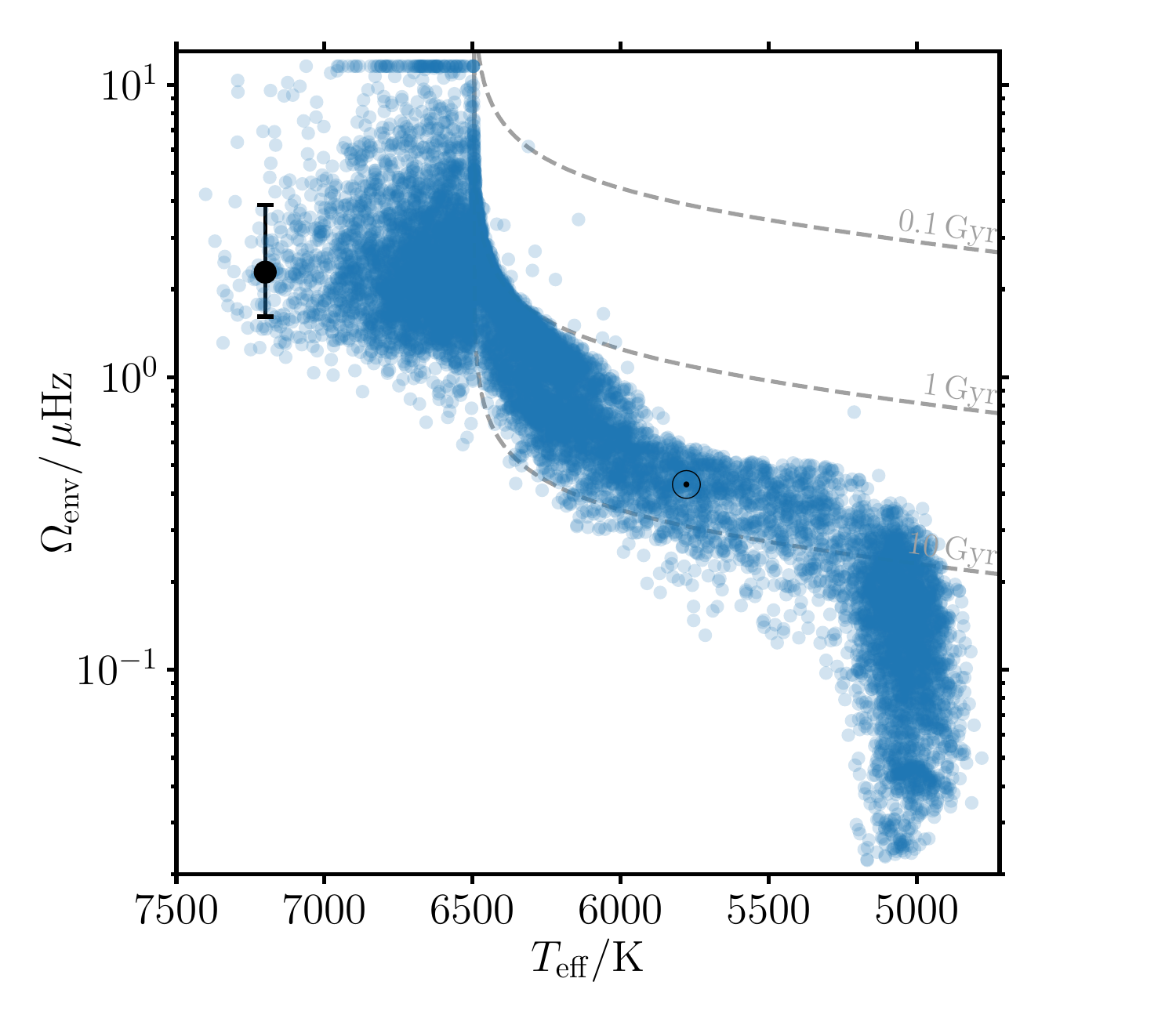}
  \caption{Envelope rotation rates for all the stars selected
      by the ATL code from the TRILEGAL simulation.  The dashed
      grey lines are rotation rates for fixed ages, from top to
      bottom, $0.1\Gyr$, $1\Gyr$ and $10\Gyr$, evaluated using the
      formula for cool stars (eq.~\ref{e:cool}).  The black point
      shows the median and standard deviation of the normal
      distribution (in period) used for stars hotter than about
      $6500\K$.
  The Sun is indicated by its usual symbol.}
  \label{f:rotation}
\end{figure}

\subsection{Rotation profiles}

A priori modelling of stellar rotation rates is an unsolved problem.
For example, stellar models broadly predict that the cores of
low-mass red giants should rotate faster than asteroseismic inferences
suggest \citep[e.g.][]{eggenberger2012,marques2013}.  Similarly, solar models
  incorporating rotation \citep[e.g.][]{turck-chieze2010} predict
  differential rotation in the Sun's radiative zone, which is at odds
  with the helioseismic inference of solid-body rotation down to about
  $0.2\Rsun$ \citep{howe2009} from the centre.  In the absence of a reliable forward
  model of stellar rotation, we used empirical relations to predict
the rotation rates of the stars in our sample.

For each star, we first compute a rotation rate predicted by the
models of \citet{angus2015} using the median values of their
parameters.  For stars with $\Teff>6500\K$, we use their formula for
hot dwarfs (their eq. 7) and draw a rotation period $P$ from a normal
distribution with mean $5.0\,\mathrm{d}$ and standard deviation
$2.1\,\mathrm{d}$.  For stars with $\Teff<6500\K$, we use
their formula for cool dwarfs (their eq. 8)
\eq{P=0.40(B-V-0.45)^{0.31}t^{0.55}\label{e:cool}}
where the colour $B-V$ is determined from $\Teff$ according to the
fitting formula by \citet{torres2010}.  We do not use the subgiant
formulae by \citet{angus2015} because their subgiant sample has stars
mostly hotter than about $6000\K$ and generally gives faster rotation
rates than observed in low-luminosity red giants.  Their model of
subgiant rotation rates was principally introduced to avoid
contaminating the gyrochronology relation for cool main-sequence
stars.  As a precaution, we set a minimum rotation period of
$1\,\mathrm{d}$ (i.e.~a maximum rotation frequency of $11.57\uHz$).

For stars on the main sequence (defined by a central hydrogen
  abundance $X_c>10^{-4}$), we assume that the star is rotating
rigidly, so that the rotation rate is constant throughout the
stellar model.  Although results for radial differential rotation in
main-sequence stars are limited, they are consistent with solid-body
rotation \citep[e.g.][]{benomar2015, nielsen2015, nielsen2017}.

For stars that have depleted hydrogen in their cores, we
  divide the rotation rate from eq.~\ref{e:cool} by
  $(R/R\st{TAMS})^2$, where $R\st{TAMS}$ is the radius of
  the star at the end of the main sequence.  This factor represents a
  na{\"i}ve conservation of angular momentum as the star expands and gives
  rotation rates that agree better with the envelope rotation rates
  found by \citet{deheuvels2014} in six subgiants and low-luminosity
  red giants.

For these post-main-sequence stars, we also
draw a core rotation rate from a normal distribution with mean
$0.375\uHz$ and standard deviation $0.105\uHz$, which is derived from
those stars in the sample studied by \citet{mosser2012rot} with large
frequency separations $\Delta\nu>12\uHz$ (i.e.~the spacing between
modes of the same degree and consecutive radial order).  If this new
rotation rate is greater than the first one, then the lower rate
is taken as rotation rate of the convective envelope
($\Omega\st{env}$) and the greater rate as the rotation rate of
the convectively-stable core ($\Omega\st{core}$).

Fig.~\ref{f:rotation} shows the envelope rotation rates of all
  the stars selected by the ATL code from the TRILEGAL simulation.

\subsection{Frequency calculation}

For each stellar model, we used
GYRE\footnote{\url{https://bitbucket.org/rhdtownsend/gyre/}}
\citep{gyre1,gyre2} to compute adiabatic mode frequencies between
$0.15$ and $0.95$ of the acoustic cut-off frequency for angular
degrees $\ell$ from $0$ to $3$.  Specifically, for all the modes, we
used a grid of 800 frequencies distributed linearly in frequency and,
for the non-radial modes, we added extra grids of 1000, 3000 and 14000
frequencies for the $\ell=1$, $2$ and $3$ modes,
distributed linearly in period.  The latter grids recover mixed modes more efficiently,
which most of our targets have.  The outer boundary condition matches
the oscillations to the oscillations in an isothermal atmosphere, as
implemented in the pulsation code ADIPLS \citep{adipls}.

Rotational splittings are computed under the standard assumptions that the
rotation is slow, is purely a function of radius, and can be treated
as a perturbation to the non-rotating mode frequencies
\citep[see e.g.][]{aerts2010}.  Under these
assumptions, the rotational splitting for a mode of radial order $n$,
angular degree $\ell$ and azimuthal order $m$ is
\eq{\delta\nu_{n\ell m}=m\beta_{n\ell}\int_0^RK_{n\ell}\Omega dr\label{e:meansplit}}
where we have defined the rotation kernels
\eq{K_{n\ell}=\frac
  {\left(\xi_r^2+(\ell(\ell+1)-1)\xi_h^2-2\xi_r\xi_h\right)r^2\rho}
  {\int_0^R\left(\xi_r^2+(\ell(\ell+1)-1)\xi_h^2-2\xi_r\xi_h\right)r^2\rho dr}}
and
\eq{\beta_{n\ell}=\frac
  {\int_0^R\left(\xi_r^2+(\ell(\ell+1)-1)\xi_h^2-2\xi_r\xi_h\right)r^2\rho}
  {\int_0^R\left(\xi_r^2+\ell(\ell+1)\xi_h^2\right)r^2\rho dr}}
In these expressions, $\xi_r$ and $\xi_h$ are the radial and
horizontal displacement eigenfunctions of the oscillation mode.
  The rotational profile $\Omega$ is given by
\eq{\Omega(r)=\left\{
\begin{array}{ll}
      \Omega\st{core} & \mathrm{if\ }r\leq r\st{BCZ} \\
      \Omega\st{env} & \mathrm{if\ }r > r\st{BCZ}
\end{array} 
\right.}
where $r\st{BCZ}$ is the radius of the base of the convective envelope.
For rigidly rotating stars
  (i.e.~$\Omega\st{core}=\Omega\st{env}$), eq.~\ref{e:meansplit} simplifies further to
\eq{\delta\nu_{n\ell m}=m\beta\Omega\st{env}}
because the rotational kernels $K_{n\ell}$ are defined to have unit integral.

\citet{davies2014b} demonstrated that a star's oscillation
  frequencies can be significantly Doppler-shifted by its
  line-of-sight velocity.  To mimic this effect in our data, we
  generated line-of-sight velocities for our stars that mimic the
  observed radial velocities for nearby stars in Gaia's second data
  release \citep[DR2,][]{gaia2018rv}.  Specifically, we fit a
  second-order polynomial in $\sin(l)$ to the mean and standard
  deviation of the radial velocities of stars with parallaxes greater
  than $0.833\,\mathrm{mas}$. This parallax corresponds to a distance
    of $1200\,\mathrm{pc}$, which would contain all but the eight most distant
    stars in our mock sample and simultaneously limits the Gaia sample to
    a relatively simple radial velocity distribution.  The selected sample
  gives the following simple
  functions of galactic longitude $l$ for the median line-of-sight
  velocity $v_r$ and its standard deviation $\sigma_{v_r}$:
\begin{align*}
  v_r(l)/\,\mathrm{km}\cdot\mathrm{s}^{-1}&=1.4+18.8\sin(l+207.7^\circ) \\
  &\qquad+ 7.9\sin(2l-5.3^\circ)\\
  \sigma_{v_r}(l)/\,\mathrm{km}\cdot\mathrm{s}^{-1}&=30.0-1.1\sin(l+260.8^\circ) \\
  &\qquad+ 4.3\sin(2l+70.6^\circ)
\end{align*}
For a star at a given galactic longitude $l$, we draw a random
  line-of-sight velocity from a normal distribution with mean $v_r(l)$
  and standard deviation $\sigma_{v_r}(l)$ and multiply the raw
  frequencies by $\sqrt{(1-v_r/c)/(1+v_r/c)}$, where $c$ is the speed
  of light.

We close this section by noting that we have not attempted to
  incorporate \emph{surface effects}: the systematic difference
  between observed and modelled mode frequencies caused by poor
  modelling of the near-surface layers of solar-like oscillators.
  Several empirical corrections have been proposed
  \citep{kjeldsen2008,ball2014,sonoi2015} and several groups have
  computed frequencies for models that incorporate information from
  three-dimensional radiation hydrodynamics simulations
  \citep[e.g.][]{sonoi2015,ball2016,jorgensen2017,trampedach2017}.
  Only a few results \citep[e.g.][]{houdek2017,sonoi2017} consider
  effects beyond the structural changes.  None of these results,
  however, is able to predict the complete surface effect for given
  stellar parameters.  As a result, we have elected not to add any
  surface effect rather than invent an empirical scheme based on the
  incomplete treatments available at this point.

\section{Lightcurve simulations}
\label{s:lightcurves}

\subsection{Introduction}

We computed artificial lightcurves using the asteroFLAG
Artificial Dataset Generator, version 3 (AADG3).  Variants of the code
have been developed over many years and extensively used, particularly
for validating the data analysis of ground-based radial velocity
measurements of solar oscillations and stellar oscillations.  The core
of the code, which simulates stochastically-driven oscillations in the
time domain, was presented by \citet{chaplin1997}.  The Solar Fitting
at Low Angular Degree Group (solarFLAG) used and developed the code to
test their data-analysis packages in two hare-and-hounds exercises
\citep{chaplin2006,jimenez2008}.  The code was further developed for
the Asteroseismic Fitting at Low Angular Degree Group
\citep[asteroFLAG, e.g.][]{chaplin2008asteroflag} from which the
current version is chiefly derived.  Details of the code were most
recently described by \citet{howe2015}.  The version used here has
been rewritten into Fortran 95 and we are now making it publicly
available\footnote{\url{https://github.com/warrickball/AADG3}} under
the GNU General Public License, version
3.\footnote{\url{https://www.gnu.org/licenses/gpl-3.0.en.html}}

The core component of AADG3 simulates the lightcurves for all modes
with the same angular degree $\ell$ and azimuthal order $m$.  The code
first generates an exponentially-damped random walk (equivalent to
a first-order autoregressive (AR) process)\footnote{See \citet{deridder2006} for an
    introduction to AR processes in the context of time-series of granulation,
    and \citet{priestley1981} or \citet{percival1993}
    for standard textbook descriptions.}
that is the same for all modes of the specified $l$ and $m$.
This is the \emph{correlated} driving term,
which we denote $u\st{c}$, and is interpreted as the component of
the granulation that contributes to exciting all modes of a given $l$
and $m$.
Then, for each radial order with the specified $l$ and $m$,
the code generates another exponentially-damped random walk, which is
the \emph{uncorrelated} driving term, denoted $u\st{u}$.  This
represents the component of the granulation that only drives a single
mode of a given $n$, $l$ and $m$.  The two sequences are added to give
an overall of driving term for that mode,
\eq{u=au\st{u} + \sqrt{1-a^2}u\st{c}\label{e:driving}}
where $a$ is a user-provided parameter, which we set at $0.45$.
\citet{toutain2006} introduced the correlated driving term to model
the asymmetry in the mode profiles.  The first $6\,\mathrm{d}$ of data
are truncated from the beginning of the sequence to allow the damped
random walk to relax into equilibrium.

To generate a complete lightcurve, AADG3 first generates the contribution
of all overtones of a given $\ell$ and $m$ using the Laplace transform
solution of a driven, damped harmonic oscillator \citep{chaplin1997}
with a sequence $u$ (eq.~\ref{e:driving}) as the driving term.
The code repeats this for each combination of $l$ and $m$.
The final lightcurve is the combination
of the lightcurves and driving terms for each $\ell$ and $m$,
weighted to reproduce the appropriate relative amplitudes
\citep[see][]{howe2015}.

\subsection{Mode lifetimes and linewidths}

Solar-like oscillations are intrinsically damped by
  near-surface convection (which also excites them) and the
  oscillations therefore have finite lifetimes.  In other words,
  the resonant peaks in
  the power spectrum, which are well-approximated by Lorentzian
  curves, have measurable linewidths (except for some very
  long-lived mixed modes in evolved solar-like oscillators).
There are currently few theoretical predictions of the linewidths of solar-like
oscillators \citep[e.g.][]{houdek2017kasc,aarslev2018} and even those
cannot be routinely and rapidly computed for a large number of
targets.  We therefore use a semi-empirical description based on data
from the nominal \emph{Kepler} mission.

We parametrise the linewidths $\Gamma$ as a function of frequency
using the same formula as \citet[][eq. 1]{appourchaux2014} and
\citet[][eq. 30]{legacy1}:
\eq{
\begin{aligned}
  \ln\Gamma
  &=\alpha\ln(\nu/\numax)+\ln\Gamma_\alpha \\
  &+\left(\frac{\ln\Delta\Gamma\st{dip}}
           {1+\left(\frac{2\ln(\nu/\nu\st{dip})}
             {\ln(W\st{dip}/\numax)}\right)^2}\right)
           \label{e:width}
\end{aligned}} %
where $\alpha$, $\Gamma_\alpha$, $\Delta\Gamma\st{dip}$,
$\nu\st{dip}$ and $W\st{dip}$ are all parameters that are
simultaneously fit as bilinear functions of $\Teff$ and $\numax$.
That is, each parameter $x$
is expressed as
\eq{x=a_x + b_x\Teff + c_x\numax\label{e:coeffs}}
where $a_x$, $b_x$ and $c_x$ are the free parameters, of which there
are 15 in total (3 for each of the 5 parameters in eq.~\ref{e:coeffs}).  We fit all 15
parameters at once, using fits to each target (with all $b_x$
and $c_x=0)$ in the LEGACY sample \citep{legacy1} as initial guesses,
to a sample containing all the radial
mode frequencies reported in the LEGACY sample as well as the 25 red
giants with highest $\numax$ in the sample studied by Davies et
al. (2018, in prep.).  Table~\ref{t:linewidths} shows the best-fitting parameters found in
this way.

\begin{table}
  \centering
  \caption{Parameters for linewidths (see eqs~\ref{e:width} and
      \ref{e:coeffs}).}
  \label{t:linewidths}
    \begin{tabular}{cccc}
      \toprule
      $x$ & $a_x$ & $b_x$ & $c_x$ \\
      \midrule
      $\alpha$ & $\sci{-3.710}{0}$ & $\sci{1.073}{-3}$ & $\sci{1.883}{-4}$ \\
      $\Gamma_\alpha$ & $\sci{-7.209}{1}$ & $\sci{1.543}{-2}$ & $\sci{9.101}{-4}$ \\
      $\Delta\Gamma\st{dip}$ & $\sci{-2.266}{-1}$ & $\sci{5.083}{-5}$ & $\sci{2.715}{-6}$ \\
      $\nu\st{dip}$ & $\sci{-2.190}{3}$ & $\sci{4.302}{-1}$ & $\sci{8.427}{-1}$ \\
      $W\st{dip}$ & $\sci{-5.639}{-1}$ & $\sci{1.138}{-4}$ & $\sci{1.312}{-4}$ \\
      \bottomrule
    \end{tabular}
\end{table}

The coupling of g- and p-modes in evolved stars affects the damping
rates \citep[see e.g.][]{basu2017}.  We divide the linewidth from
equation~\ref{e:width} by the ratio $Q_{n\ell}$
\eq{Q_{n\ell}=\frac{\mathcal{I}_{n\ell}}{\mathcal{I}_0(\nu_{n\ell})}\label{e:Qnl}}
where $\nu_{n\ell}$ and $\mathcal{I}_{n\ell}$ are the frequency and
inertia of the mode with radial order $n$ and angular degree
$\ell$, and $\mathcal{I}_0(\nu)$ is the mode inertia of the radial
modes interpolated at the frequency $\nu$.  Because mixed modes have
greater inertiae than pure p-modes, $Q_{n\ell}$ is greater than one,
so the mixed modes have narrower linewidths (i.e.~they live longer)
than the pure p-modes.

\subsection{Mode amplitudes}

To predict the intrinsic mode amplitudes of the stellar oscillations in the power spectrum,
we follow the prescription by \citet{chaplin2011}, which is itself based on
results by \citet{kjeldsen1995} and \citet{samadi2007}.

We assume that the maximum rms amplitude of the radial modes can be
scaled from the solar value by
\eq{A^\mathrm{rms}\st{max}=A^\mathrm{rms}_{\mathrm{max},\odot}\beta
  \left(\frac{L}{\Lsun}\right)\left(\frac{M}{\Msun}\right)^{-1}
  \left(\frac{\Teff}{T_{\mathrm{eff},\odot}}\right)^{-2}\label{e:Ascale}}
where $A^\mathrm{rms}_{\mathrm{max},\odot}=2.1\,\mathrm{ppm}$ in the
TESS bandpass and we have taken $T_{\mathrm{eff},\odot}=5777\K$.
Besides the factor $\beta$, this is in essence the scaling relation
  presented by \citet{kjeldsen1995}.  Compared to the amplitudes measured
  by \citet{legacy1} for the LEGACY sample (i.e.~dwarfs observed
  for at least one year during the nominal \emph{Kepler} mission),
  eq.~(\ref{e:Ascale}) is consistent within about 40 per cent,
  and within 25 per cent for all but 6 of the 66 stars in the sample.

The factor $\beta$ is defined by
\eq{\beta=1-\mathrm{exp}\left(\frac{\Teff-T\st{red}}{\Delta T}\right)}
and corrects the formula for the apparent decrease in the amplitudes
of the hottest dwarfs.  Here, $\Delta T=1250\K$ and $T\st{red}$
is the temperature of the red edge of the $\delta$-Scuti instability
strip at the star's luminosity, which we take as
\eq{T\st{red}=8907\K\cdot\left(\frac{L}{\Lsun}\right)^{-0.093}}
For the envelope's full-width half-maximum (FWHM) width
$\Gamma\st{env}$, we use the scaling relation \citep{mosser2012}
\eq{\Gamma\st{env}=0.66\uHz\cdot\left(\frac{\numax}{\uHz}\right)^{0.88}}
If $\Teff > T_{\mathrm{eff},\odot}$, we multiply $\Gamma\st{env}$
by the factor $1+\sci{6}{-4}(\Teff-T_{\mathrm{eff},\odot})$ (Lund et al., in prep.).
The rms power in the mode with radial order $n$ and angular degree
$\ell$ is then
\eq{\left(A^\mathrm{rms}_{n\ell}\right)^2=\left(A^\mathrm{rms}\st{max}\right)^2
  \exp\left[-\frac{(\nu_{n\ell}-\numax)^2}{2\sigma\st{env}^2}\right]}
where $\sigma\st{env}=\Gamma\st{env}/2\sqrt{2\ln 2}$.

Like the linewidths, the mode powers are also affected by the coupling
of p- and g-modes.
We divide the mode powers by $Q_{n\ell}$ (see eq.~\ref{e:Qnl}) so that
the more strongly coupled modes are suppressed.  Finally, to
  avoid simulating lightcurves for modes that contribute negligibly
  to the power spectrum, we restrict the list of modes to those with
  heights in the power spectrum greater than $10^{-4}$ times the
  expected granulation background signal.

\subsection{Background properties}

To determine the characteristic timescale of the granulation
$\tau\st{gran}$, we use equation 10 of \citet{kjeldsen2011},
\eq{\tau\st{gran}
  =\left(\frac{\nu\st{max}}{\nu_{\mathrm{max},\odot}}\right)^{-1}{\tau_{\mathrm{gran},\odot}}}
with
a solar value $\tau_{\mathrm{gran},\odot}=250\s$.  The amplitude of
the granulation is given by combining equations 24 and 21 of
\citet{kjeldsen2011}, which gives
\eq{\sigma\st{gran}\propto\frac{L^2}{M^3\Teff^{5.5}}\numax}
The granulation signal drives the oscillations.  After it is generated
for the mode calculation, it is added to the lightcurve after an
appropriate scaling \citep[see][]{howe2015}.
Any other background processes that are not correlated with the
  oscillations (e.g. supergranulation) can be added to the lightcurves
  we provide.  The granulation timeseries is generated by averaging over
  50 subcadences of each cadence, which also apodises the signal.
  The oscillations are not apodised.

Because we simulate the granulation with a first-order
  autoregressive process, its power spectrum is \citep[e.g.][]{deridder2006}
\eq{P\st{gran}(\nu)=\frac{4\sigma\st{gran}^2\tau\st{gran}}{1+(2\pi\nu\tau\st{gran})^2}}
which is often referred to as a Harvey law \citep{harvey1985}.
  While the use of a fixed power $2$ in the denominator is reasonable,
  many studies leave the power free and often fit data better with
  powers around $4$ \citep[e.g.][]{michel2009, kallinger2014}.  Our
  results are limited by the need to simulate the
  granulation signal
  in the time domain (rather than the frequency domain).  While we aim
  to improve on the background model, or at least provide more freedom
  in how it is modelled (e.g.~higher-order autoregressive models),
  we regard this as beyond the scope of the current work.

The amplitude of the
white noise is generated according to the same formulae as used in the
ATL code.  The noise model is inferred from the target's $I$-band
  magnitude, which is similar to the expected magnitude in the TESS
bandpass.

\subsection{Mode visibilities}

The apparent amplitudes of the modes are influenced by two main
  geometric effects: cancellation and inclination.  First, as the
  angular degree $\ell$ increases, there are more and more
  equally-sized brighter and darker regions across the stellar
  surface, which cancel out when integrated over the visible stellar surface.
  This cancellation can be quantified by a visibility $V_\ell$, often
  normalised by the radial mode's visibility $V_0$ to give the
  normalised visibilities $\tilde{V}_\ell\equiv V_\ell/V_0$.  The
  power of a non-radial mode of degree $\ell$ is then multiplied by
  $\tilde{V}_\ell$.

  Though these visibilities can in
  principle be computed theoretically \citep[e.g.][]{ballot2011vis},
  \citet{legacy1} found that the predictions disagreed with
  observations, especially for the $\ell=3$ modes.  We have opted to
  use the median normalised visibilities for the main-sequence stars
  studied by \citet{legacy1}, which are $\tilde{V}_1=1.505$,
  $\tilde{V}_2=0.620$ and $\tilde{V}_3=0.075$.
  \citet{legacy1} found no significant correlations with the stars'
  properties and found values in reasonable agreement with the red
  giants studied by \citet{mosser2012}.
  These mode visibilities are sufficiently realistic for our purposes
    but the visibilities do depend on the photometric bandpass and will
    be different in the actual TESS data.

Second, a rotating star's inclination angle influences the
  relative visibility of modes of different azimuthal order $m$.
  Assuming equipartition of energy between each of the $2\ell+1$
  components of a rotationally split multiplet,
  the power in each $m$ component is the intrinsic mode
  power multiplied by a factor \citep{gizon2003}
\eq{\mathcal{E}_{\ell m}(i)=\frac{(\ell-|m|)!}{(\ell+|m|)!}
  \left[P_\ell^{|m|}(\cos i)\right]^2}
where $i$ is the inclination angle of the rotation axis
  and $P_\ell^{|m|}$ is an associated Legendre polynomial.
  
We assumed that the rotation axis of each star is randomly distributed
over the sphere (i.e.~it can point in any direction), which implies
that $\cos i$ is uniformly distributed.  We assign the
inclination $i$ for each star by drawing a uniform variate
$w\sim U(0,1)$ and assigning $i=\cos^{-1}w$.

\subsection{Other parameters}

Finally, we summarize our choices for the remaining global parameters
in AADG3.  As mentioned before, the sequence of driving terms is
allowed to relax for $6\,\mathrm{d}$, which, given the 2-minute
cadence of the TESS data, corresponds to $4320$ data points.  The
output lightcurves contain $255418$ or $257345$ points, corresponding to the
total length of all the sectors in either the northern or southern
ecliptic hemispheres.  The full lightcurves are subsequently divided
into lightcurves for each sector in which a given star is observed.

\section{Catalogue contents}
\label{s:catalogue}

\subsection{Stellar properties and lightcurves}

The lightcurves are publicly available as archives for each
sector.\footnote{\dataset[\!\!\doi{10.5281/zenodo.1470155}]{\doi{10.5281/zenodo.1470155}}}
  The data repository also includes a table of comma-separated values
  (CSV) containing the data from all of the headers for each
  lightcurve, for quick analysis of the sample and target selection.
We also separately provide all the scripts that
  were used to produce and manipulate the
  data.\footnote{\url{https://github.com/warrickball/s4tess}}

Each
lightcurve is a FITS file with a filename of the form
\begin{center}
  \texttt{<ID>\_<SECTOR>\_<SEC\_RANK>.fits}
\end{center}
where
\begin{itemize}
  \item \texttt{<ID>} is the overall rank of the star in the sample,
    which identifies it uniquely;
  \item \texttt{<SECTOR>} is the sector number, from 1 to 26
    (inclusive); and
  \item \texttt{<SEC\_RANK>} is the rank assigned by the ATL code for that
    sector.
\end{itemize}
So, for example, \texttt{00123\_17\_050.fits} would be the
lightcurve for the 123rd star in the sample when observed in
the fourth sector in the northern ecliptic hemisphere (the
  seventeenth sector overall), in which the ATL ranked it 50th
for the detectability of its oscillations.
The file \texttt{00123\_18\_056.fits} would be a lightcurve
  for the same star when observed in the fifth sector in the northern
  ecliptic hemisphere, in which it was ranked 56th for the
  detectability of its oscillations.

Each file contains a header and two arrays of data, with details
  of each component given in Table~\ref{t:file}.  The header contains
  a number of overall properties of the star, taken from TRILEGAL
  simulation, the ATL results, the MESA stellar models and the input
  for AADG3.  The first array contains the lightcurve data.  The
  second array contains the mode frequency information used by AADG3
  to create the lightcurve.  Thus, each lightcurve file
  contains all the information required to recreate the AADG3 input
  and the public pipeline repository includes a script to do this.

\begin{table*}
  \caption{Detailed contents of the lightcurve FITS files.  Each row
    gives a key's name, its units (if applicable) and a short
    description.}
  \begin{center}
    \begin{minipage}{\textwidth}
\begin{tabular}[t]{rcp{4.8cm}}
\toprule
\multicolumn{3}{c}{Header (\texttt{PRIMARY})} \\
Key & Unit & Description \\
\midrule
\texttt{ID} & & rank in whole sample \\
\texttt{SECTOR} & & TESS observing sector \\
\texttt{SEC\_RANK} & & rank in this sector \\
\texttt{TOT\_RANK} & & rank in whole sky (including stars not observed by TESS) \\
\texttt{PMIX} & & detection probability \\
\texttt{MASS} & $\Msun$ & stellar mass ($M$) \\
\texttt{RADIUS} & $\Rsun$ & stellar radius \\
\texttt{AGE} & $\mathrm{Gyr}$ & stellar age \\
\texttt{TEFF} & $\K$ & effective temperature ($\Teff$) \\
\texttt{LOGG} & $\mathrm{cm}\cdot\mathrm{s}^{-2}$ & $\log_{10}$ of surface gravity ($\log g$) \\
\texttt{LUM} & $\Lsun$ & stellar luminosity ($L$) \\
\texttt{X\_C} & & central hydrogen abundance \\
\texttt{Y\_C} & & central helium abundance \\
\texttt{Z\_INI} & & initial metal abundance \\
\texttt{FE\_H} & & final metallicity ($\FeH$) \\
\texttt{DELTA\_NU} & $\uHz$ & large separation ($\Delta\nu$, from scaling relations) \\
\texttt{NU\_MAX} & $\uHz$ & frequency of maximum oscillation power ($\numax$, from scaling relations) \\
\texttt{BETA} & & red edge amplitude correction factor ($\beta$) \\
\texttt{A\_RMSMAX} & $\mathrm{ppm}$ & maximum rms power of radial modes ($A^\mathrm{rms}\st{max}$) \\
\texttt{GAMMA\_ENV} & $\uHz$ & FWHM of oscillation power envelope ($\Gamma\st{env}$) \\

\texttt{OMEGA\_C} & $\uHz$ & central/core rotation rate ($\Omega\st{core}$) \\
\texttt{OMEGA\_E} & $\uHz$ & surface/envelope rotation rate ($\Omega\st{env}$) \\
\texttt{VR} & $\mathrm{km}\cdot\mathrm{s}^{-1}$ & radial velocity ($v_r$) \\
\texttt{MU0} & & distance modulus \\
\texttt{AV} & & interstellar reddening \\
\bottomrule
\end{tabular}
\begin{tabular}[t]{rcp{4.8cm}}
  \toprule
  \multicolumn{3}{c}{Header (\texttt{PRIMARY}, cont.)} \\
Key & Unit & Description \\
\midrule
\texttt{ELON} & $\degrees$ & ecliptic longitude \\
\texttt{ELAT} & $\degrees$ & ecliptic latitude \\
\texttt{GLON} & $\degrees$ & galactic longitude \\
\texttt{GLAT} & $\degrees$ & galactic latitude \\
\texttt{GC} & & galactic component: 1, 2, 3 or 4 for the thin disc, thick disc, halo or bulge \\
\texttt{COMP} & & binarity: 0 if the star is single or 1 or 2 if the star is the primary or secondary in a binary \\
\texttt{SIGMA} & $\mathrm{ppm}$ & white noise amplitude \\
\texttt{SEED} & & seed for random number generator \\
\texttt{N\_CADS} & & number of cadences in hemisphere \\
\texttt{GRAN\_SIG} & $\mathrm{ppm}$ & granulation amplitude ($\sigma\st{gran}$) \\
\texttt{GRAN\_TAU} & $\mathrm{s}$ & granulation timescale ($\tau\st{gran}$) \\
\texttt{INC} & $\degrees$ & inclination angle \\
\\
\midrule
\multicolumn{3}{c}{Lightcurve array (\texttt{LIGHTCURVE})} \\
Key & Unit & Description \\
\midrule
\texttt{TIME} & MJD & days since first observation \\
\texttt{FLUX} & $\mathrm{ppm}$ & fractional intensity variation \\
\texttt{CADENCENO} & & cadences since first observation \\
\\
\midrule
\multicolumn{3}{c}{Mode data (\texttt{MODES})} \\
Key & Unit & Description \\
\midrule
\texttt{L} & & angular degree ($\ell$) \\
\texttt{N} & & radial order ($n$) \\
\texttt{FREQ} & $\uHz$ & frequency ($\nu_{n\ell}$) \\
\texttt{WIDTH} & $\uHz$ & linewidth ($\Gamma_{n\ell}$) \\
\texttt{POWER} & $\mathrm{ppm}^2$ & RMS power \\
\texttt{ROT} & $\uHz$ & rotation splitting ($\delta\nu_{n\ell0}$) \\
\bottomrule
\end{tabular}
\end{minipage}

  \end{center}
  \label{t:file}
\end{table*}

\subsection{Example power spectra}

\begin{figure}
  \centering
  \includegraphics[width=\columnwidth]{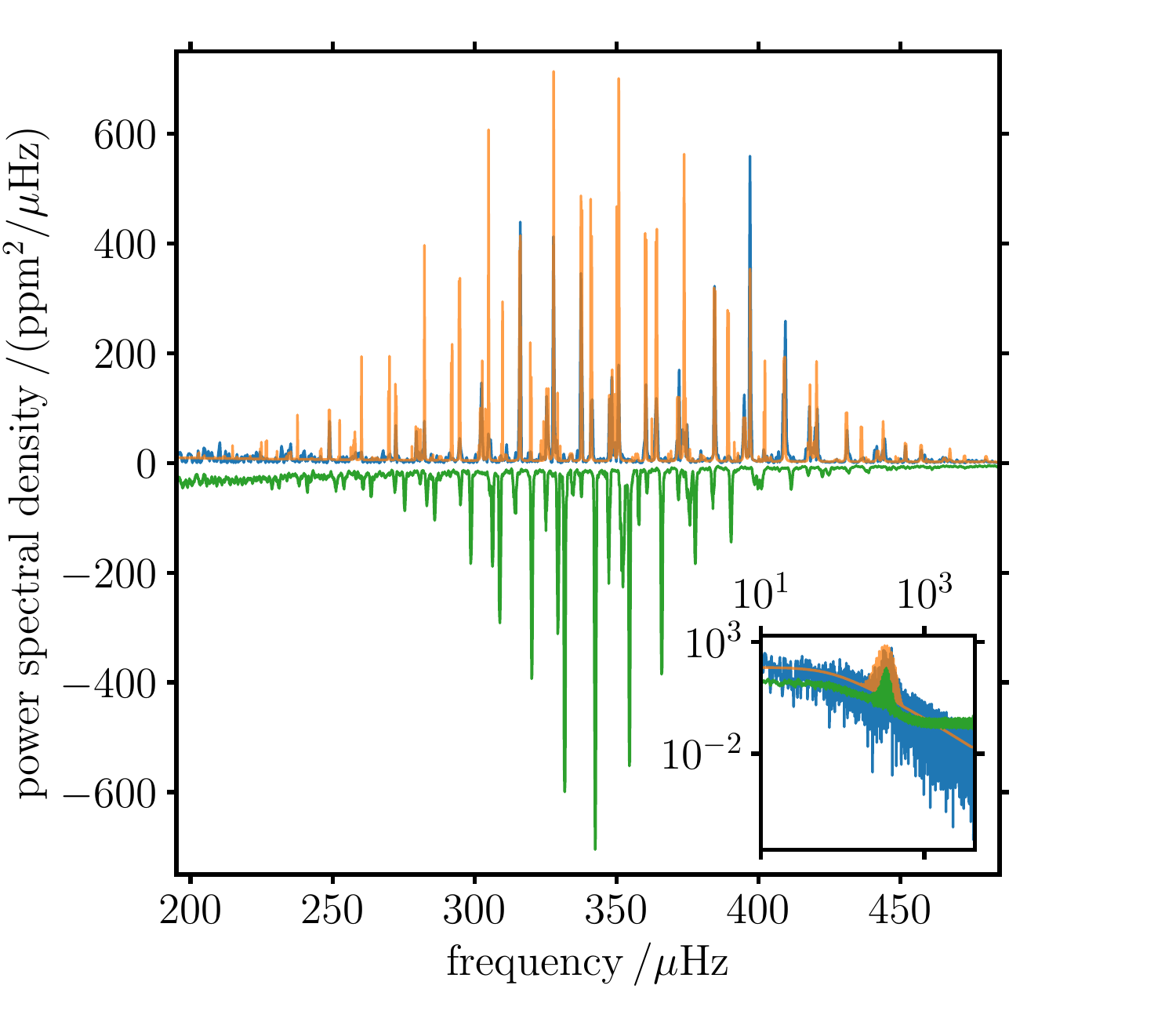}
  \caption{Power spectrum of a typical target (star 00197 observed in
    sector 2) in the range where the oscillations are clearest.
    The orange line is the mean spectral model with no mode
      asymmetry.  The green line is a reflection of a 50-point
      boxcar-smoothed power spectrum of KIC~8179973.
      The inset shows the complete power spectra,
        with the \emph{Kepler} power spectrum divided by 10 for clarity.}
  \label{f:typical}
\end{figure}

\begin{figure}
  \centering
  \includegraphics[width=\columnwidth]{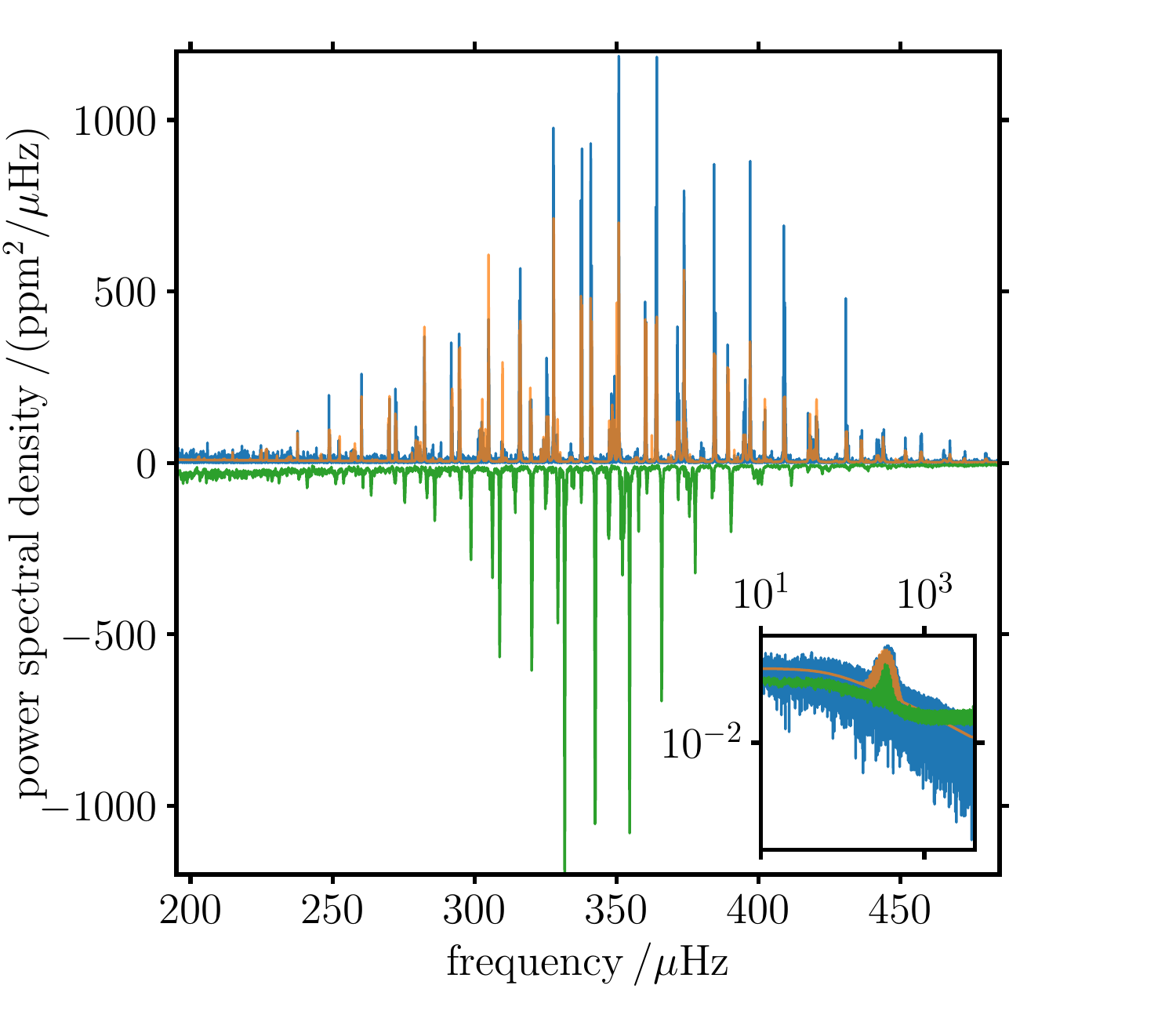}
  \caption{Power spectrum of a typical target (star 00197, as in
    Fig.~\ref{f:typical}) observed for the maximum duration in the
    southern hemisphere.
    The orange line is the mean spectral model with no mode
      asymmetry.  The green line is a reflection of a 20-point
      boxcar-smoothed power spectrum of KIC~8179973.
    The inset shows the complete power spectra,
        with the \emph{Kepler} power spectrum divided by 10 for clarity.}
  \label{f:full}
\end{figure}

\subsubsection{Typical low-luminosity red giant}

As expected from the ATL and can be seen in Fig.~\ref{f:kiel}, most of
the best targets in our sample are found at the base of the red giant
branch, with $\Teff\approx5000\K$ and $\log g\approx3.5$.  This is
mainly an effect of the relationship between mode amplitude and
luminosity.  We would in principle prioritise stars further up the red
giant branch too but the oscillations of these stars will be available
from the full-frame images, with a cadence of 30 minutes.  The ATL
therefore places a lower limit on $\numax$ of $240\uHz$.

Fig.~\ref{f:typical} shows the power spectrum of star 00197,
  located near the southern ecliptic pole, as observed in sector 2.
With $\Teff=4933\K$ and $\log g=3.47$, this star is typical of the
bulk of targets on the lower red giant branch.  Fig.~\ref{f:full}
shows a power spectrum of the same star but this time computed
from the full, roughly year-long lightcurve.  For comparison,
  we have also included the power spectrum of the similar star
  KIC~8179973 ($\Teff\approx4949\K$, $\log g\approx3.48$)
  which was observed by \emph{Kepler} during its nominal mission.
  We computed this and other \emph{Kepler} power spectra from concatenated
  timeseries prepared for asteroseismic analysis by \citet{handberg2014}.

\begin{figure}
  \centering
  \includegraphics[width=\columnwidth]{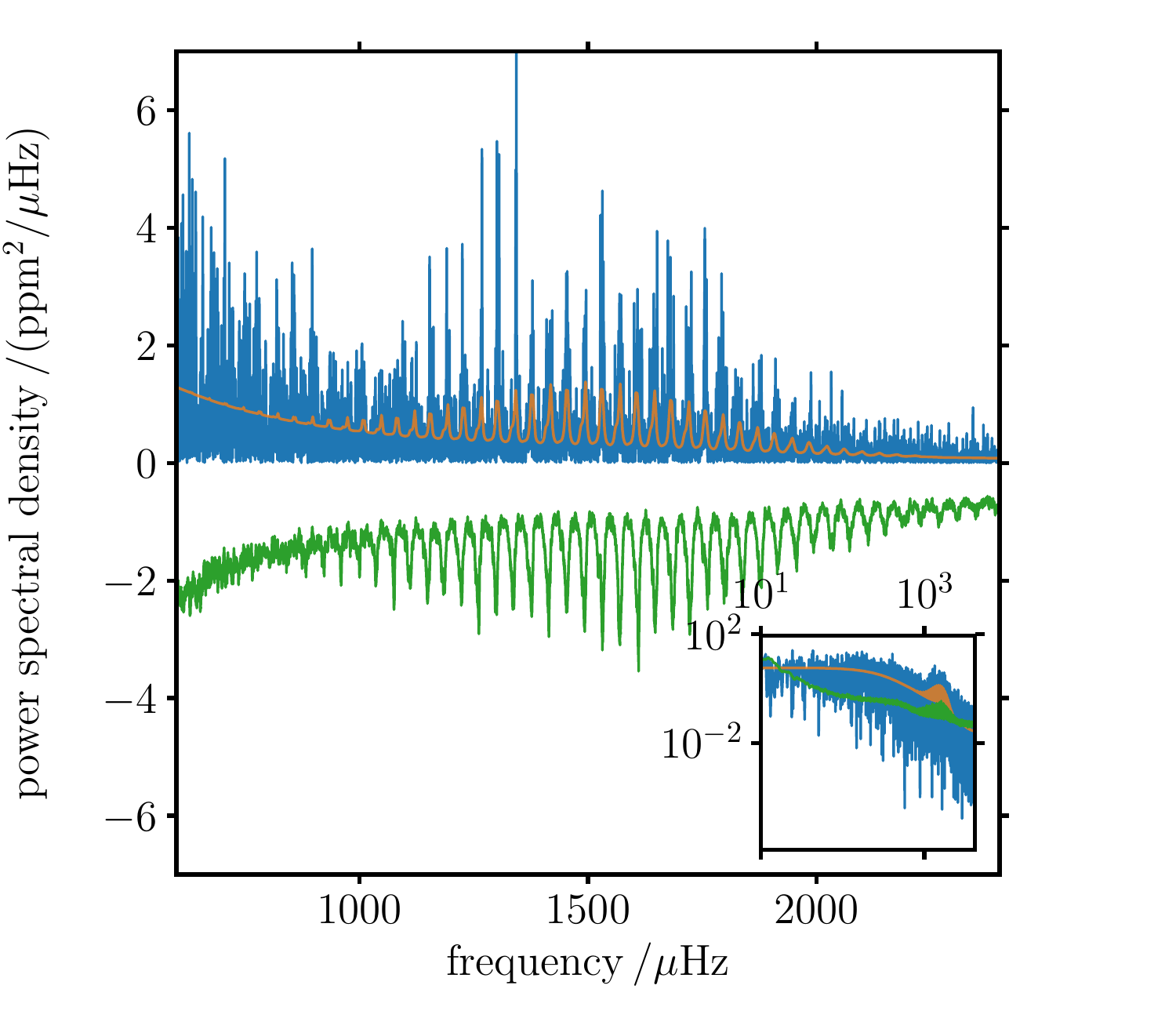}
  \caption{Power spectrum for one sector of data of an F-type dwarf,
    similar to KIC~11253226 (star 00704).  The orange line is the mean
    spectral model with no mode asymmetry. The green line is a
      reflection of a 200-point boxcar-smoothed power spectrum of
      KIC~11253226. The inset shows the complete power spectra,
        with the \emph{Kepler} power spectrum divided by 10 for clarity.}
  \label{f:Ftype}
\end{figure}

\begin{figure}
  \centering
  \includegraphics[width=\columnwidth]{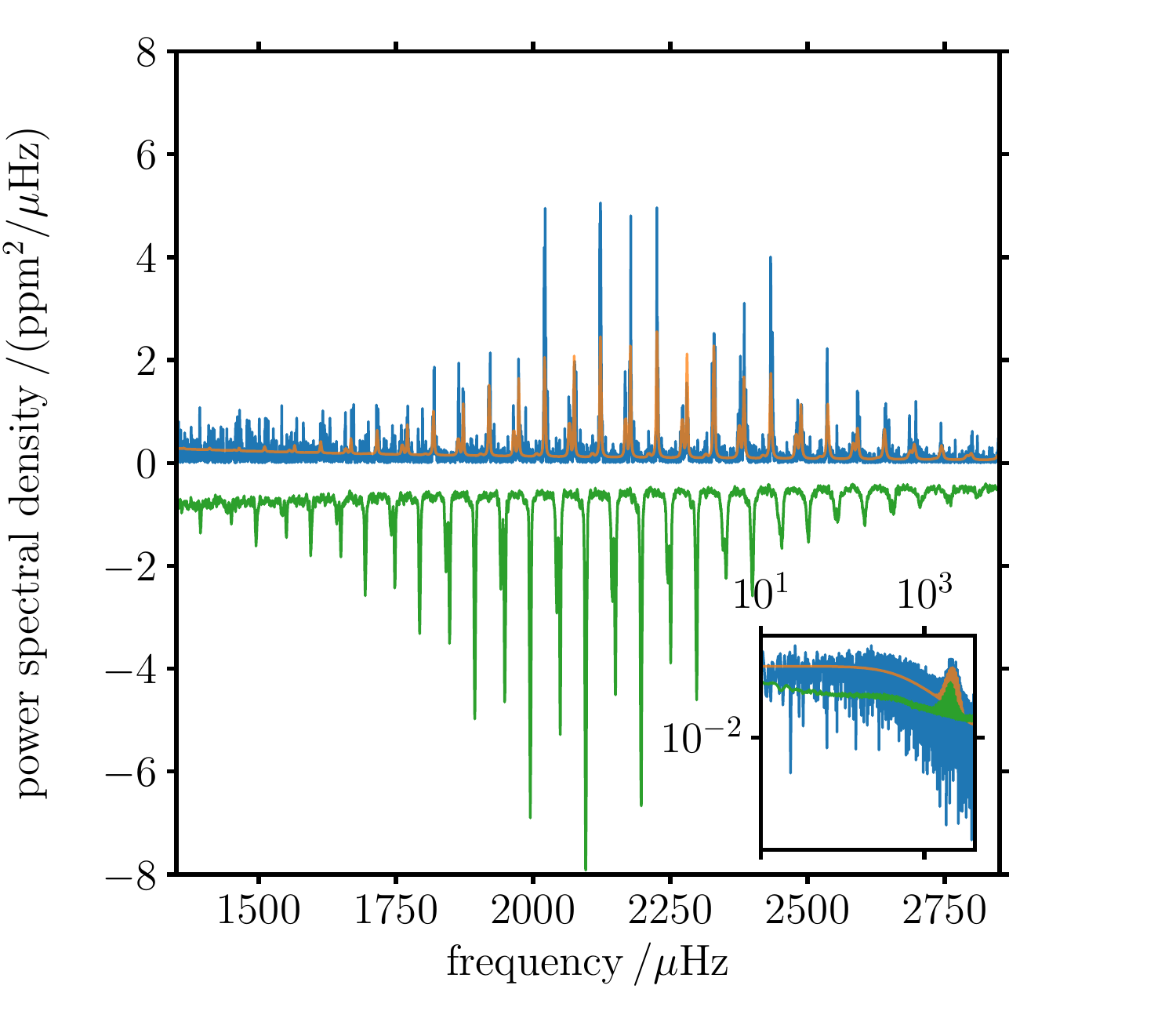}
  \caption{Power spectrum for one sector of data of a G-type dwarf,
    similar to KIC~6106415 and KIC~6116048 (star 00771).  The orange
    line is the mean spectral model with no mode asymmetry.
    The green line is a reflection of a 200-point
      boxcar-smoothed power spectrum of KIC~6116048.  The
      inset shows the complete power spectra,
        with the \emph{Kepler} power spectrum divided by 10 for clarity.}
  \label{f:Gtype}
\end{figure}

\subsubsection{Main-sequence stars}

Although the sample is dominated by subgiants and low-luminosity red giants, there
are also less evolved stars.  Fig.~\ref{f:Ftype} shows the power
spectrum of star 00704, observed in sector 25.  This star has similar
properties ($\Teff=6636\K$, $\log g=4.17$) to the known solar-like
oscillator KIC~11253226 ($\Teff\approx6642\K$, $\log g\approx4.18$), which
\citet{legacy1} and \citet{legacy2} studied as part of the
\emph{Kepler} LEGACY sample.  A 200-point boxcar-smoothed power
  spectrum for KIC~11253226 is shown in green.

Fig.~\ref{f:Gtype} shows the power spectrum of star 00771, observed in
sector 18.  This star is more Sun-like ($\Teff=6052\K$, $\log g=4.31$)
and similar to the \emph{Kepler} targets KIC~6106415 ($\Teff\approx6037\K$,
$\log g\approx4.31$) and KIC~6116048 ($\Teff\approx6033\K$, $\log g\approx4.29$), which
were also part of the LEGACY sample.  A 200-point boxcar-smoothed power
  spectrum for KIC~6116048 is shown in green.  Though the $\ell=0$ modes are
clearly distorted by the $\ell=2$ modes, it remains to be seen whether
or not the frequencies can be disentangled reliably.

\section{Discussion}
\label{s:discussion}

The present simulated results offer reasonably realistic predictions
of what the underlying signal from solar-like oscillators observed by
TESS will look like.  Our lightcurves are pristine, in the sense
  that they contain no noise, though we provide the expected white noise
  parameters in the lightcurve files.
We have not considered other effects that would degrade the
power spectra, including instrumental effects or observing gaps.  We
have also not included other stellar signals like starspots or
transits.  Any of these, however, can be straightforwardly added to
the lightcurves we provide.

A more complicated contaminant is the effect of frequency changes
caused by magnetic activity.  As the frequencies vary over the course
of an observation, so the mode profiles in the power spectrum are
broadened and potentially biased \citep{chaplin2008magbias}.  The
  effect is probably small in most TESS targets because the roughly
  month-long observations are much shorter than known activity cycles
  \citep[e.g.][]{borosaikia2018}.  AADG3 has inherited the
  capability of modelling these frequency shifts from earlier
versions of the code \citep{howe2015} but it has not yet been
validated in the new version of the code released with this article.
In addition, this would require a further semi-empirical model with
which to predict the activity cycle periods and the magnitude of the
frequency shifts for each star.

Finally, the code is still limited by our limited ability to predict
relevant oscillation parameters, in particular the rotation
profiles of the stars.  The linewidths (or damping rates) also
  require an empirical model but at least they are well constrained by
  observations from \emph{Kepler}.  These are areas of active
research and continued asteroseismic analyses will provided important
constraints on theoretical models.

Our methods are naturally applicable to any set of timeseries
observations of solar-like oscillators, be it preparation for upcoming
missions like PLATO \citep{plato}, planning for ongoing projects like
the Stellar Oscillations Network Group \citep[SONG,][]{song} or
testing new analyses of existing datasets like \emph{Kepler} or COROT.
The PLATO consortium already operates a lightcurve generator, the
PLATO Solar-like Light-curve
Simulator\footnote{\url{https://sites.lesia.obspm.fr/psls/}} (PSLS),
which is based on the COROT simulator
\citep[\texttt{simuLC},][]{baudin2007}.  The PLATO and COROT
simulators produce an oscillation lightcurve by the inverse Fourier
transform of a model Fourier spectrum and our method (which works
entirely in the time-domain) is complementary.

\section{Conclusions}
\label{s:conclusions}

We have presented a catalogue of mock observations of solar-like
oscillators observed by NASA's TESS mission in its short-cadence mode.
Our artificial data combines a simulation of Milky Way
populations, detailed stellar models and empirical relations for less
well understood physical processes.  Targets have been selected from
the galaxy simulation using the same method as has been used for the
actual mission and the sample therefore reflects the same selection
effects.  Together, these provide realistic lightcurves with which to
prepare for the steady stream of data expected from TESS.

Our artificial lightcurves are publicly available\footnote{\dataset[\!\!\doi{10.5281/zenodo.1470155}]{\doi{10.5281/zenodo.1470155}}} and can be extended to
include various phenomena that we have excluded, be they instrumental
effects or other astrophysical signals.  The methods we have presented
are also applicable to any observing programme for solar-like
oscillations and will be useful in the future for observatories like
SONG and PLATO.

\acknowledgements

WHB would like to thank Tom Barclay for providing detailed help with
and information about TESS's orbit and pointings.  AM
  acknowledges support from the European Research Council (ERC) under
  the European Union's Horizon 2020 research and innovation programme
  (project ASTEROCHRONOMETRY, grant agreement no. 772293).  The
authors thank the UK Science and Technology Facilities Council (STFC)
for support under grant ST/R0023297/1.
AM, GRD, and LG are grateful to the International Space Science
  Institute (ISSI) for support provided to the asteroSTEP ISSI
  International Team.
Calculations in this paper
made use of the University of Birmingham's BlueBEAR High-Performance
Computing service.\footnote{\url{http://www.birmingham.ac.uk/bear}}
Previous development of AADG3 has been supported by the International Space
Science Institute (ISSI), through a workshop programme award, and by
the European Helio- and Asteroseismology Network (HELAS), a major
international collaboration funded by the European Commission’s Sixth
Framework Programme.

\software{NumPy\footnote{\url{http://www.numpy.org}} \citep{numpy},
  SciPy\footnote{\url{http://www.scipy.org}} \citep{scipy},
  Astropy\footnote{\url{http://www.astropy.org}} \citep{astropy1, astropy2},
  Matplotlib\footnote{\url{http://matplotlib.org}} \citep{matplotlib},
  MESA\footnote{\url{http://mesa.sourceforge.net}} \citep{paxton2011,paxton2013,paxton2015},
  TRILEGAL \citep{girardi2005},
  ADIPLS \citep{adipls}
  GYRE \citep{gyre1, gyre2},
  AADG3 \citep{aadg3.0.0}
}

\bibliographystyle{aasjournal}
\bibliography{../master}

\end{document}